\documentclass[prl,twocolumn,superscriptaddress,10pt]{revtex4-1}
\usepackage{amsmath,amssymb}
\usepackage[svgnames]{xcolor}
\usepackage{graphicx}
\usepackage{natbib}

\usepackage{mathtools}
\DeclarePairedDelimiter\abs{\lvert}{\rvert}
\renewcommand{\vec}[1]{\mathbf{#1}}
\renewcommand{\section}[1]{\textit{#1}.---}

\usepackage{acro}
\DeclareAcronym{les}{
  short=LES,
  long=Large Eddy Simulations,
  long-format=\itshape,
}
\DeclareAcronym{rans}{
  short=RANS,
  long=Reynolds Average Navier-Stokes,
  long-format=\itshape,
}
\DeclareAcronym{dns}{
  short=DNS,
  long=Direct Numerical Simulation,
}
\DeclareAcronym{ghost}{
  short=GHOST,
  long=Geophysical High-Order Suite for Turbulence,
  long-format=\itshape,
}
\DeclareAcronym{wexac}{
  short=WEXAC,
  long=Weizmann Exascale Cluster,
  long-format=\itshape,
}

\DeclareAcronym{dss}{
  short=DSS,
  long=Direct Statistical Simulation,
  long-format=\itshape,
}
\DeclareAcronym{ce}{
  short=CE,
  long=Cumulant Expansion,
  long-format=\itshape,
}
\DeclareAcronym{ssst}{
  short=SSST,
  long=Stochastic Structural Stability Theory,
  long-format=\itshape,
}

\usepackage[colorlinks=true,linkcolor=DarkBlue,urlcolor=DarkBlue,citecolor=FireBrick]{hyperref}

\begin{document}

\title{Turbulence statistics in a 2D vortex condensate}

\author{Anna Frishman}
\email{frishman@princeton.edu}
\affiliation{Princeton Center for Theoretical Science, Princeton University, Princeton, New Jersey 08544, USA}
\affiliation{Department of Physics of Complex Systems, Weizmann Institute of Science, P.O. Box 26, 76100 Rehovot, Israel}
\author{Corentin Herbert}
\email{corentin.herbert@ens-lyon.fr}
\affiliation{Univ Lyon, ENS de Lyon, Univ Claude Bernard, CNRS, Laboratoire de Physique, F-69342 Lyon, France}
\affiliation{Department of Physics of Complex Systems, Weizmann Institute of Science, P.O. Box 26, 76100 Rehovot, Israel}

\begin{abstract}
  Disentangling the evolution of a coherent mean-flow and turbulent fluctuations, interacting through the non-linearity of the Navier-Stokes equations, is a central issue in fluid mechanics.
  It affects a wide range of flows, such as planetary atmospheres, plasmas or wall-bounded flows, and hampers turbulence models.
  We consider the special case of a two-dimensional flow in a periodic box, for which the mean-flow, a pair of box-size vortices called \emph{condensate}, emerges from turbulence through an inverse cascade process.
  As was recently shown, a perturbative closure describes correctly the condensate when turbulence is excited at small scales.
  In this context, we obtain explicit results for the statistics of turbulence, encoded in the Reynolds stress tensor.
  We demonstrate that the two components of the Reynolds stress, the momentum flux and the turbulent energy, are determined by different mechanisms.
  It was suggested previously that the momentum flux is fixed by a balance between forcing and mean-flow advection: using unprecedently long numerical simulations, we provide the first direct evidence supporting this prediction.
  By contrast, combining analytical computations with numerical simulations, we show that the turbulent energy is determined only by mean-flow advection, and obtain for the first time a formula describing its profile in the vortex.
\end{abstract}


\maketitle


More often than not, turbulence appears hand in hand with a coherent mean-flow.
Yet, an understanding of the interactions between the two remains elusive.
At high Reynolds numbers, these interactions are strong, and closed equations describing seperately the mean-flow and the turbulent fluctuations cannot be obtained.
This is a central problem in fluid mechanics, with far-reaching consequences both for fundamental physics and engineering~\cite{TennekesLumleyBook,*TownsendBook,*PopeBook}.
The Reynolds stress tensor is a key object for mean-flow/turbulence interactions, encapsulating the important physical quantities: the turbulent momentum flux and the turbulent energy.
It is through the Reynolds stress that turbulence retroacts on the mean-flow.
Thus, numerical models which do not fully resolve turbulence, such as \acl{les} or \acl{rans}, require a parametrization of the Reynolds stress~\cite{Spalart2000,*LesieurLESBook}.
It is therefore crucial to better understand the statistics of this object.

Recently, significant progress in this direction was achieved in the realm of 2D flows~\cite{Falkovich2016}.
In 2D, turbulence can spontaneously generate a mean-flow, rather than feeding on it.
Indeed, the turbulent energy tends to be transferred towards increasingly larger scales, in what is termed an \emph{inverse cascade}~\cite{Kraichnan1980,*Boffetta2012}.
Eventually, if the large-scale dissipation mechanism is slow enough, energy accumulates at the domain scale and a mean-flow emerges, referred to as a \emph{condensate}~\cite{Sommeria1986,*LSmith1993,*LSmith1994,*Chertkov2007,*Xia2009}.
Its structure depends on the domain geometry: in the following, we focus on the vortex condensate, which appears in a square box or on a flat torus (as part of a vortex dipole).
The condensate is expected to become asymptotically strong compared to turbulence when the large-scale dissipation rate tends to zero, which justifies a perturbative treatment.
More precisely, if the mean-flow shear rate is much larger than the rate of non-linear turbulence-turbulence interactions, then the latter can generically be neglected.
This is called the quasi-linear approach; it was used extensively to study dynamics in numerical simulations~\cite{Schneider2006b,*OGorman2007,*Srinivasan2012}, often under different names such as \ac{dss}, \ac{ce}~\cite{Tobias2013,*Marston2016,*AitChaalal2016} or \ac{ssst}~\cite{Farrell2003,*Farrell2007}, and justified theoretically using \emph{adiabatic reduction}~\cite{Bouchet2013}.
Actually, if turbulence is excited at asymptotically small scales, the perturbative treatment allows to analytically derive an explicit formula for the mean-flow and momentum flux profiles, as demonstrated for the vortex condensate~\cite{Laurie2014,Kolokolov2016a}, and discussed for jets~\cite{Woillez2017,Frishman2017b} and on the sphere~\cite{Falkovich2016}.
Until now, the only part of these predictions which was quantitatively checked against data from \ac{dns} is the profile of the mean-flow~\cite{Laurie2014}.

In this Letter, we present new results on the statistics of the Reynolds stress.
First, using long time integration, we provide the first numerical evidence supporting the explicit formula for the momentum flux~\cite{Laurie2014,Kolokolov2016a}.
Second, we show that the turbulent energy is determined by a different mechanism; we explain its structure by combining a first-principles theoretical framework and numerical results, hence describing the full Reynolds tensor.
To our knowledge, this is the first time that such explicit formulas are derived either in 2D or 3D turbulence.
Furthermore, while the results for the mean-flow and momentum flux rely on features specific to 2D turbulence, the mechanism governing turbulent energy may apply more generally.

\section{Framework and numerical methods}
We consider an incompressible flow on a square domain of length $L$ with periodic boundary conditions, with linear friction as the large-scale dissipation mechanism.
The governing equations for the velocity field $\vec{v}$ are the 2D Navier-Stokes equations:
\begin{equation}
\partial_t \vec{v} + \vec{v} \cdot \nabla \vec{v} = - \nabla P -\alpha \vec{v} - \nu (-\Delta)^{p/2} \vec{v} + \vec{F}, \label{eq:ns}
\end{equation}
where $P$ is the pressure, $\alpha$ the friction coefficient, $\nu$ the hyperviscosity, and $\vec{F}$ a random forcing.
We work with an isotropic, white in time forcing acting in a narrow shell in Fourier space centered on wave number $K_f$, with $\varepsilon=\langle \vec{v}\cdot \vec{F}\rangle$ the average energy injection rate.

\begin{figure}
  \includegraphics[width=0.8\linewidth]{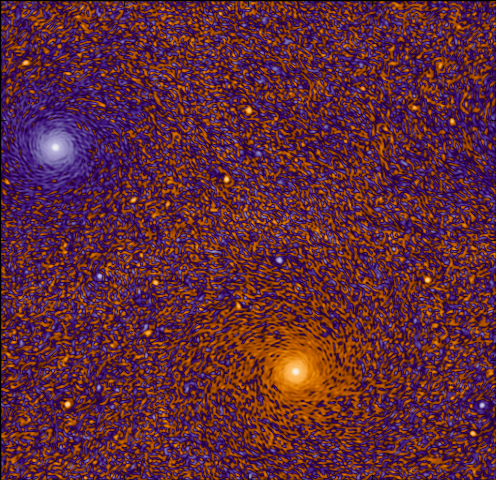}
  \caption{\label{fig:snapshot} (color online) Snapshot of the vorticity field (1024\textsuperscript{2}) in the stationary state of the \ac{dns}; the vortex dipole forming the condensate is clearly visible.}
\end{figure}

\ac{dns} results are obtained by integrating~\eqref{eq:ns} using the \ac{ghost} pseudo-spectral code, at resolution 512\textsuperscript{2} and 1024\textsuperscript{2}, and with parameters $L=2\pi, K_f=L/\ell_f=100, p=16, \nu=5\cdot 10^{-35}$.
\begin{table}
  \begin{tabular}{ccccccc}
    \hline
    \hline
                         & Z    & A    & B    & C   & D    & E \\
    \hline
    $10^5 \times \alpha$ & 20   & 11   & 5.5  & 2.5 & 1.25 & 0.625\\
    $10^3 \times \delta$ & 12.3 & 8.14 & 4.58 & 2.4 & 1.45 & 1.09\\
    \hline
    \hline
  \end{tabular}
  \caption{\label{tab:params} Parameters for the \ac{dns} runs.}
\end{table}
We carried out runs covering almost two orders of magnitude in the friction coefficient $\alpha$ (see Table~\ref{tab:params}).
In agreement with previous studies, the flow reaches a \emph{condensate} steady-state, taking the form of a vortex dipole (see Fig.~\ref{fig:snapshot}).
The vortices drift slowly across the box, with fast turbulent fluctuations superimposed onto them.
We carry out a Reynolds decomposition in polar coordinates centered on one of the vortices: the mean-flow $\langle \vec{v} \rangle= U\vec{e}_\phi$ is purely azimuthal, while the fluctuations read $\tilde{\vec{v}}=v\vec{e}_r+u\vec{e}_\phi$.
Angular brackets $\langle \cdot\rangle$ denote a time average~\footnote{The averaging is performed in the reference frame of the vortex, which is equivalent to an average over times shorter than the vortex motion.}.
The dimensionless parameter $\delta = \alpha L^{2/3}/\varepsilon^{1/3}$, introduced as the ratio between the condensate spin-down time $\alpha^{-1}$ and the eddy turnover time $L^{2/3}\varepsilon^{-1/3}$, measures the timescale separation.

As shown previously~\cite{Laurie2014,Kolokolov2016a,Frishman2017b}, within the quasi-linear approximation --- justified for $\delta \ll 1$ --- and once $K_f \gg1$, so that the mean flow can be approximated locally by a uniform shear, it is possible to derive an explicit formula for the mean-flow $U$ and momentum flux $\langle uv \rangle$:
\begin{equation}
  U = \sqrt{3\varepsilon/\alpha}, \qquad \langle uv \rangle = -r \sqrt{\alpha \varepsilon/3}.
  \label{eq:Uuv}
\end{equation}
This profile is expected to hold in the range $\ell_f \ll r \ll R_u$, where $R_u=\delta^{-1/2} K_f^{-2/3} L$ measures the radius where the rate of non-linear interactions is comparable to the mean-flow shear rate.
At larger radii the quasi-linear approximation is expected to break, while for $r\ll \ell_f$ the uniform shear approximation for the mean-flow is no longer applicable.
Of course, once $R_u \gtrsim L$ the range of validity should be set by the boundary --- be it a wall or a second vortex.
The mean-flow profile in our simulations (shown in Supplementary Material \cite{Suppmaterial}) is compatible with the theoretical prediction in a region that expands with decreasing $\delta$.
This is consistent with earlier numerical results~\cite{Laurie2014}, over a wider range of friction coefficient $\alpha$.
In the figures, we represent the range $\ell_f \ll r \ll R_u$ by a shaded area and the empirical range of validity by vertical dashed lines.
Focusing on this region, we now discuss the three terms of the average Reynolds stress tensor, $\langle uv \rangle$, $\langle u^2\rangle$ and $\langle v^2 \rangle$.

\section{The average momentum flux profile}
\begin{figure}
  \includegraphics[width=\linewidth]{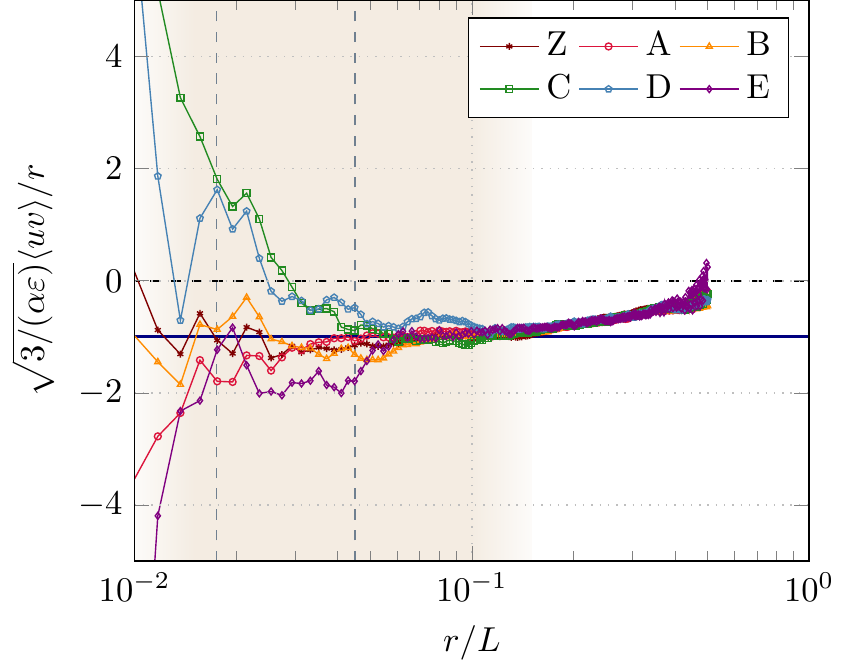}
  \caption{\label{fig:uv} (color online) Rescaled profile of average momentum flux $\langle uv \rangle$ from \ac{dns}. The horizontal blue line corresponds to the theoretical prediction. Vertical dashed lines denote the range where the mean flow profile~\eqref{eq:Uuv} is observed to hold.}
\end{figure}
Until now, unlike the mean-flow, no experimental or numerical evidence has been given for the average momentum flux~\eqref{eq:Uuv}.
The difficulty is that it is a small quantity --- $\langle uv \rangle/U^2=O(\delta^{3/2})$ --- and it is not sign definite: the average value results from cancellations of strongly fluctuating contributions.
Using the hybrid parallelization of \ac{ghost}~\cite{Mininni2011b}, we were able to integrate our runs over extremely long times (about 320 000 turnover times) and accumulate enough statistics to observe partial convergence of the average momentum flux $\langle uv \rangle$, shown in Fig.~\ref{fig:uv}.
Although convergence is restricted to a sub-region of the region of interest which does not match the empirical range defined above, we find that the momentum flux is consistent with the prediction~\eqref{eq:Uuv} in that region: the numerical data unambiguously confirms the negative sign and the scaling with $\alpha$, and is compatible with the theoretical prediction at the outer edge of the region, $r/L\lesssim 10^{-1}$ (smaller radii have less statistics).

\section{The average turbulent energy profile}
Symmetry considerations imply that the magnitude of the diagonal terms of the Reynolds tensor, $\langle u^2\rangle$ and $\langle v^2\rangle$, as well as the mechanism that dictates them, are different than those for the momentum flux~\cite{Laurie2014,Kolokolov2016a}.
We therefore take a different approach than~\cite{Kolokolov2016b}, where the turbulent energy is computed within the same framework used for the momentum flux.
Instead, we will show that $\langle u^2\rangle$ and $\langle v^2\rangle$ are determined by the zero modes of the advection equation for two-point correlation functions.
Indeed, in the quasi-linear framework, fluctuations are linearly advected by the mean-flow.
Due to incompressibility, the dynamics is characterized by a single field:  vorticity, for instance, obeys the equation $(\partial_t+\mathcal{D}) \omega + L_U \omega = f_\omega$, with $\mathcal{D}=\alpha+\nu(-\Delta)^{p/2}$, and $L_U\omega=\frac{U}{r}\partial_\phi \omega + \Omega' v$.
%
From here, a closed equation for two-point correlation functions directly follows; for $\Phi=\langle \omega_1 \omega_2 \rangle$, with $C=\langle f_{\omega 1}f_{\omega 2} \rangle$, the steady-state advection equation reads $\lbrack L_U^{(1)}+L_U^{(2)\dagger}+\mathcal{D}_1+\mathcal{D}_2^\dagger \rbrack \Phi = 2C$.
In general, the operator $L_U$ is non-local and this is an integro-differential equation (mixed $\langle \omega_1 v_2 \rangle$ terms appear).
Besides, we are eventually interested in velocity statistics.
Therefore, we transform this equation into a partial differential equation for the radial velocity correlation function, using incompressibility: $\partial_{\phi_1}\partial_{\phi_2}\Phi=\Delta_1\Delta_2 r_1r_2\langle v_1v_2 \rangle$.
We now claim that $\langle v_1 v_2\rangle$ is dominated by zero modes of the advection operator.
This entails two approximations.
First, we neglect the contribution from dissipation.
This is justifiable in the region $r_1,r_2\ll R_u$, where fluctuations are weak compared to the mean-flow.
In addition, we neglect the injection by the forcing, which is justified as long as $\abs{r_1-r_2}\gg \ell_f$.
In fact, we eventually need to take the limit $r_1\to r_2$, passing through $\ell_f$, to compute the energy $\langle u^2\rangle$ and $\langle v^2\rangle$.
However, for $K_f\gg1$, we expect to recover the correct result in this limit, since velocity correlation functions are expected to be continuous passing through this point --- the energy being mainly determined at scales larger than the forcing scale.
We shall solve explicitly the resulting \emph{homogeneous Lyapunov equation} characterizing the zero modes, with the mean-flow profile~\eqref{eq:Uuv}.
It reads
\begin{equation}
\left[ \mathcal{L}_{2}r_2 \left( 2 \partial_{r_1}r_1 + \mathcal{L}_{1}\right) - \mathcal{L}_{1}r_1 \left( 2 \partial_{r_2}r_2 + \mathcal{L}_{2}\right)\right]\partial_{\phi_1} \langle v_1 v_2\rangle = 0,
\label{eq:fluct}
\end{equation}
with the notation $\mathcal{L}_{i}=r_i^2 \Delta_i$, using isotropy ($\partial_{\phi_1}=-\partial_{\phi_2}$).

Given the form of the advection operator, it is natural to decompose the fluctuations into angular harmonics: $v(r,\phi) = \sum_{m=-\infty}^{\infty} \hat{v}_m(r) e^{im\phi}$ (and similarly for $u$).
The resulting equation for $\langle \hat{v}_m(r_1) \hat{v}_m^*(r_2)\rangle$ is scale invariant, and independent of the value of the mean flow $U$.
This prompts the use of the scaling form $\langle \hat{v}_m(r_1)\hat{v}^*_m(r_2)\rangle= r_1^{\lambda}f_m(r_2/r_1)$ for the solution, which allows to convert our PDE into an ODE in the variable $R=r_2/r_1$.
It can then be written compactly in terms of $R$ and $\bar{\lambda}=\lambda+1$ as the Hypergeometric equation:
\begin{equation}
\prod_{i=1}^4\left(R\frac{d}{dR}-\gamma_i\right)f_m(R)=R\prod_{i=1}^4\left(R\frac{d}{dR}+\alpha_i\right)f_m(R),
\label{eq:hyprg}
\end{equation}
with the parameters $(\gamma_1,\gamma_2,\gamma_3,\gamma_4)=(\bar{\lambda}-m,\bar{\lambda}+m,-1+\sqrt{m^2-1},-1-\sqrt{m^2-1})$, $(\alpha_1,\alpha_2,\alpha_3,\alpha_4)=(1-m,1+m,-\bar{\lambda}+\sqrt{m^2-1},-\bar{\lambda}-\sqrt{m^2-1})$.
This equation has four families of solutions, described in details in the Supplementary Material~\cite{Suppmaterial}, each parameterized by $m$ and $\lambda$.
We now turn to the numerical simulations to identify which solutions derived in our theoretical framework (i.e. which parameters $\lambda$) contribute to the turbulent energy profile.

\begin{figure}
  \includegraphics[width=\linewidth]{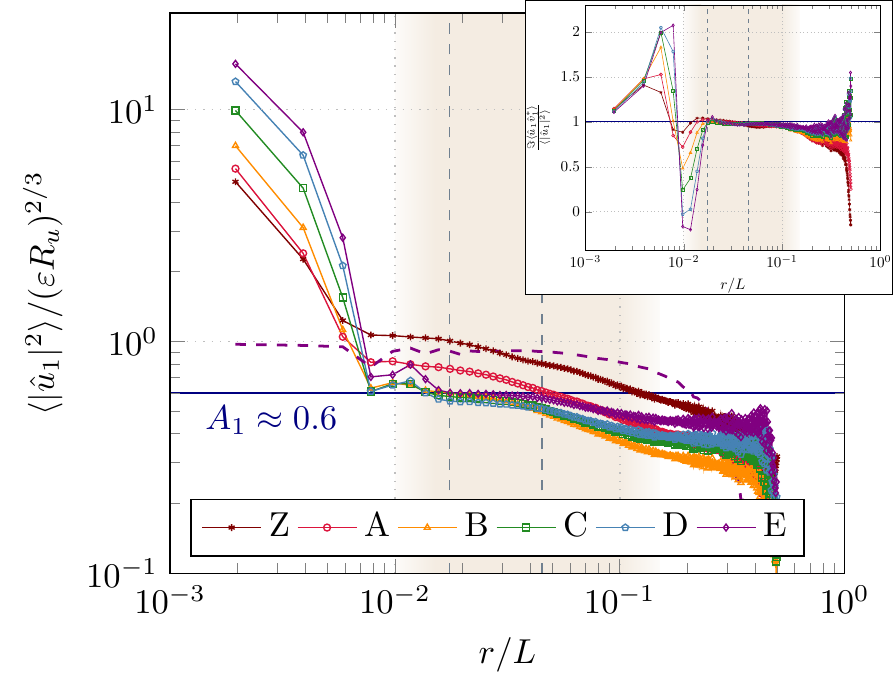}
  \caption{\label{fig:harm1} (color online) Profile of the first harmonic $\langle |\hat{u}_1|^2 \rangle$ for all the runs, rescaled according to~\eqref{eq:harmon1}. The dashed purple line corresponds to the same quantity, normalized by the full field $\langle u^2\rangle/2$ for run E. Inset: the ratio  $\Im\langle \hat{u}_1\hat{v}_1^*\rangle /\langle \abs{\hat{u}_1}^2 \rangle$.}
\end{figure}

\begin{figure}
  \includegraphics[width=\linewidth]{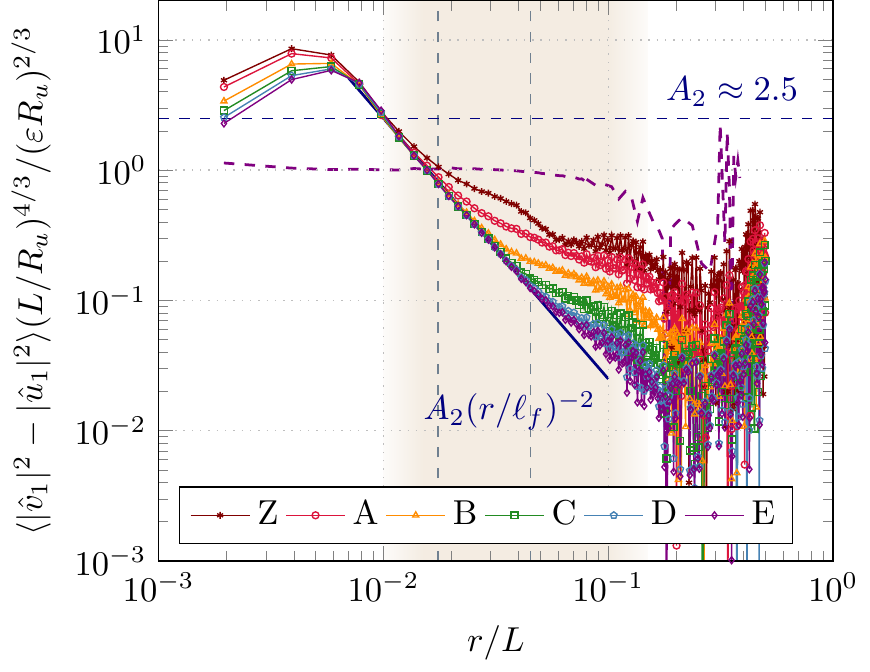}
  \caption{\label{fig:harm1diff} (color online) Rescaled profile of the power law part for the first harmonic $\langle |\hat{v}_1|^2 - |\hat{u}_1|^2 \rangle$, for all the runs. The dashed purple line corresponds to the same quantity, normalized by $\langle v^2-u^2\rangle/2$ for run E.}
\end{figure}

Decomposing the \ac{dns} data into harmonics, we see that turbulent energy in the region of interest is strongly dominated (about 90\%) by the $m=1$ modes (see the dashed curve in Fig.~\ref{fig:harm1} and~\ref{fig:harm1diff}, and also Supplementary Material~\cite{Suppmaterial}).
Figure~\ref{fig:harm1} also suggests that, in the universal region, $\langle |\hat{u}_{1}|^2\rangle\approx \Im\langle\hat{u}_{1}\hat{v}^*_{1}\rangle\approx \text{Const}$ and $\langle |\hat{v}_{1}|^2\rangle\approx \text{Const}+r^{\beta}$ with some $\beta<0$.
Comparing this form to the possible solutions to~\eqref{eq:hyprg}, we find that it corresponds to a unique superposition of two solutions, one with $\lambda=0$, which has  $\langle |\hat{u}_{1}|^2\rangle=\langle |\hat{v}_{1}|^2\rangle$, and the second with $\lambda=\beta=-2$, \cite{Suppmaterial}.
Indeed, a very good match to such a power law can be seen in Fig.~\ref{fig:harm1diff}, once $\langle |\hat{u}_{1}|^2\rangle$ is subtracted from $\langle |\hat{v}_{1}|^2\rangle$.
In its current form, our theory does not determine the scaling of the correlation functions with $\delta$.
We again turn to the \ac{dns} results, and find  that $\langle |\hat{u}_{1}|^2\rangle$ and $\Im\langle\hat{u}_{1}\hat{v}^*_{1}\rangle$ collapse for the different runs when rescaled by  $\delta^{-1/3}$, while the $r^{-2}$ part of $\langle |\hat{v}_{1}|^2\rangle$ collapses under rescaling by $\delta^{-1}$.
However, we also have $U^2\propto \delta^{-1}$ and we have relied throughout on the ratio $\langle |\hat{v}_1|^2\rangle/U^2$ being small.
Thus, we expect this term to be suppressed by some power of $K_f^{-1}$.
We finally arrive at the form
\begin{align}
  \langle \abs{\hat{v}_1}^2 \rangle &=  (\varepsilon R_u)^{2/3}\left[A_1+A_2 \left(\frac{R_u}{L}\right)^{4/3}\left(\frac{\ell_f}{r}\right)^2\right],\nonumber\\
    \langle \abs{\hat{u}_1}^2 \rangle &= A_1(\varepsilon R_u)^{2/3}, \quad \langle \hat{u}_1 \hat{v}_1^*\rangle = iA_1(\varepsilon R_u)^{2/3}, \label{eq:harmon1}
\end{align}
 where the dependence on $K_f$ is a plausible guess, which gives $A_1$ and $A_2$ as order one numerical coefficients.
In this case, we have that at most $\langle |\hat{v}_1|^2\rangle/U^2 \sim K_f^{-4/3}$, which is obtained at $r=\ell_f$.
Equation~\eqref{eq:harmon1} is our main quantitative result, since it provides a formula for the turbulent energy $\langle u^2\rangle$ and $\langle v^2\rangle$.
The agreement with these formulas observed in Fig.~\ref{fig:harm1} and~\ref{fig:harm1diff} improves with decreasing $\delta$.

While it is too computationally expensive to measure two-point correlation functions in \ac{dns}, our analytical solution does contain this information.
In particular, the contribution of the first harmonic to the second order velocity increment can be written quite compactly:
\begin{equation}
\begin{split}
\frac{\langle (\vec{v_1}-\vec{v_2})^2\rangle}2 =4A_1(\varepsilon R_u)^{2/3}(1-\cos\Delta\phi) \\+2A_2(\varepsilon L)^{2/3} \left(\frac{R_u}{L}\right)^2 \frac{\ell_f^2}{r_1 r_2}\frac{(\vec{r_1}-\vec{r_2})^2}{r_1 r_2}+..
\end{split}
\end{equation}

The mode $m=1$, which dominates the energy, is the lowest mode that is determined by mean-flow advection.
Indeed, there is no interaction between the $m=0$ harmonic and the mean-flow, as a consequence of the latter being isotropic.
We also remark that the zeroth harmonic of the radial velocity, $v$, identically vanishes due to incompressibility.
Formulas for higher-order harmonics, however, can be deduced from our formalism.
Let us demonstrate that with $m=2$.
\begin{figure}
  \includegraphics[width=\linewidth]{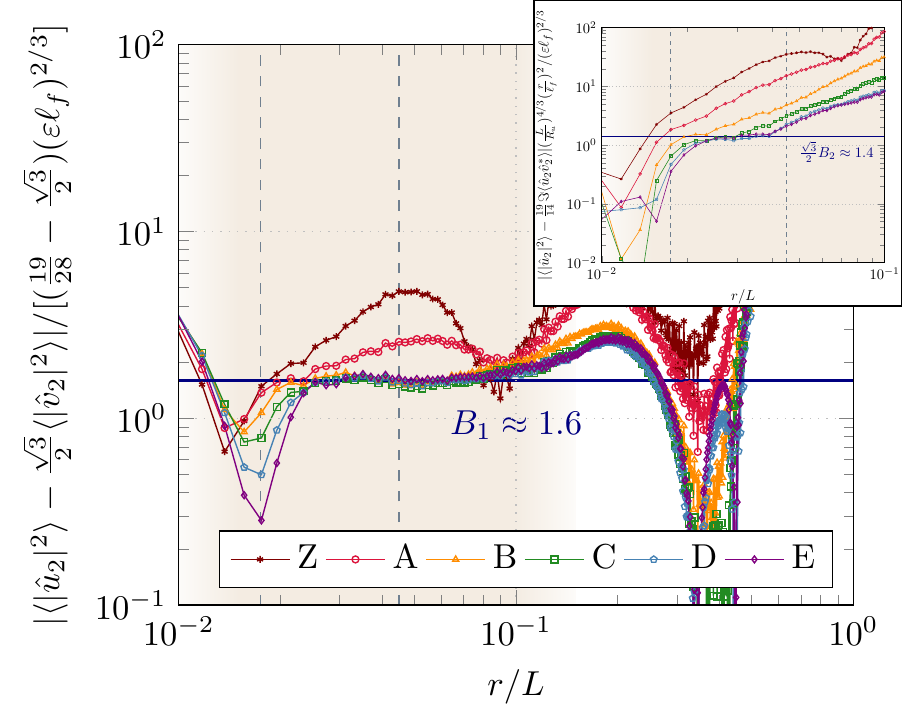}
  \caption{\label{fig:harm2} (color online) Combination of the second harmonics demonstrating the constant and $r^{-2}$ (inset) contributions to the profiles~\eqref{eq:harmon2}.}
\end{figure}
The \ac{dns} results indicate that $\langle \hat{u}_2 \hat{v}_2^*\rangle$ is constant in the universal region.
In addition, since $\lambda=-2$ contributes to $m=1$, it seems reasonable to check if it can contribute to $m=2$ as well.
This leads to a superposition of three solutions: the first has $\lambda=0$, and is of the form $f^{(1)}_2(R)=R^{\sqrt{3}-1}g(R)$, for $R\leq1$, where $g(R)$ is a polynomial of second degree.
The two other solutions have $\lambda=-2$, and they come in a particular combination which reads $f^{(2)}_2(R)=R^{\sqrt{3}-1}- (\sqrt{3}/2) R $ for $R<1$ \cite{Suppmaterial}.
Using the \ac{dns} data, we make a plausible guess for the coefficients in this combination of solutions by identifying the scaling with $\delta$ and requiring an order one magnitude.
This gives the energy in the second angular harmonic:
\begin{align}
  \langle \abs{\hat{v}_2}^2 \rangle &=  (\varepsilon \ell_f)^{2/3} \left\lbrack B_1 -B_2 \left(\frac{R_u}{L}\right)^{4/3} \left(\frac{\ell_f}{r}\right)^2\right\rbrack,\nonumber \\
    \langle \abs{\hat{u}_2}^2 \rangle &= (\varepsilon \ell_f)^{2/3} \left\lbrack \frac{19}{28} B_1 -\frac{\sqrt{3}}2 B_2 \left(\frac{R_u}{L}\right)^{4/3} \left(\frac{\ell_f}{r}\right)^2\right\rbrack,\nonumber \\
     \langle \hat{u}_2 \hat{v}_2^*\rangle &= i\frac{B_1}{2} (\varepsilon \ell_f)^{2/3}. \label{eq:harmon2}
\end{align}
The form~\eqref{eq:harmon2} seems to be compatible with \ac{dns} data, as shown in Fig.~\ref{fig:harm2}, except close to the inner boundary of the region, which could be either due to the contribution of another solution or an effect of the forcing.

We finally note that an analog of~\eqref{eq:fluct}  was used in a previous attempt~\cite{Chertkov2010} to determine the momentum flux and mean-flow.
There, the hierarchy was closed at the level of cubic terms.
As was later realized~\cite{Laurie2014,Kolokolov2016a}, this produces a zero momentum flux at any order, because the equations are invariant under the transformation $\phi\to -\phi$, $t\to-t$, while the momentum flux breaks this symmetry.
The same happens with the zero modes discussed here: because the coefficients in~\eqref{eq:hyprg} are real, so are the solutions $\langle \hat{v}_m(r_1)\hat{v}^*_m(r_2) \rangle$, resulting in $\Re \lbrack \langle \hat{u}_m\hat{v}^*_m\rangle \rbrack=0$.

\section{Conclusion}
In this Letter, we have explored the turbulent fluctuations statistics in a 2D vortex mean-flow sustained by turbulence.
We demonstrated that the turbulent energy and momentum flux are governed by different mechanisms.
Through a combination of \ac{dns} and first-principles theoretical analysis, we showed, for the first time, that the turbulent energy profile~\eqref{eq:harmon1} --- and, more generally, two-point correlation functions --- is determined by zero modes of the mean-flow advection equation.
The contribution of these zero modes to the momentum flux vanishes, which explains why it is determined at next order by a balance between forcing and advection.
We provided the first evidence supporting the resulting profile.
A consequence is that the turbulent energy and momentum flux scale differently with the small parameter $\delta$, at variance with the assumption used to justify the quasi-linear approach in a kinetic theory framework~\cite{Bouchet2013}.
Moreover, while the limit $K_f\gg1$ was discussed before~\cite{Woillez2017}, our results point to its crucial role in suppressing the fluctuations, which was not considered.
We have relied upon \ac{dns} results to identify which theoretical solutions are realized and their scaling with $\delta$.
This leaves open the question of their selection mechanism.
In particular, it would be interesting to check if in a box with solid boundaries, such as found in experiments, the turbulent energy profile would be the same, or if instead other solutions would be selected.
Our results also represent an important first step for understanding turbulence statistics in more complex flows, such as geophysical jets, plasmas, and ultimately, long-standing problems such as turbulent boundary layers.

\begin{acknowledgments}
  Computer time was provided by the Weizmann Institute of Science on the \ac{wexac}. 
  Authors are listed in alphabetical order and contributed equally to this work.
\end{acknowledgments}

\bibliographystyle{apsrev4-1}
\bibliography{bibtexlib}

\end{document}